# Restricting the h-index to a citation time window: A case study of a timed Hirsch index

**Michael Schreiber**

*Institute of Physics, Chemnitz University of Technology, 09107 Chemnitz, Germany.*
*Phone: +49 371 531 21910, Fax: +49 371 531 21919*
*E-mail: schreiber@physik.tu-chemnitz.de*

The h-index has been shown to increase in many cases mostly because of citations to rather old publications. This inertia can be circumvented by restricting the evaluation to a citation time window. Here I report results of an empirical study analyzing the evolution of the thus defined timed h-index in dependence on the length of the citation time window.

## 1. Introduction

A recently popular measure of scientific achievements of a researcher is given by the Hirsch index or h-index as defined by Hirsch (2005). It can easily be determined after sorting the publications of a researcher in decreasing order according to their citation frequencies. The index is then given by the largest rank $h$ which is equal to or smaller than the number of citations to the $h$-th paper. It is nowadays often used as a performance indicator for evaluation purposes and fund allocations. Beside principle questions about the usefulness of single number indicators for such purposes there occurs a specific problem when the h-index is applied in such away. This is due to the fact that the index cannot decrease and that its increase is often dominated by citations to *rather old* publications. This inert behavior of the index (Schreiber, 2013) has raised doubts whether the h-index is a reasonable measure for predicting future scientific impact. Specifically, observed high correlations between h-index values of researchers in early and late stages of their careers (Hirsch, 2007) can be explained as structural correlations due to an order restriction (García-Pérez & Núñez-Antón, 2013), because the h-index cannot decrease.

A solution could be the restriction of the determination of the h-index to *recent* publications in order to measure the recent performance of a researcher. I have recently analyzed citation records in such away, namely by discarding older publications and taking only papers published in the last years into account (Schreiber, 2014). In the present investigation I analyze my own citation data in a retrospective way, namely determining the evolution of the thus restrictively defined timed h-index $h_t$ in dependence on the length $t$ of the utilized citation time window. These proceedings allow me to obtain insights into the changing impact of my publications during my career.

For the same purpose, Pan & Fortunato (2014) have proposed the Author Impact Factor which extends the Journal Impact Factor to individual authors. Using a 5-year publication window, they described trends and variations of the impact in terms of citations in the 6[th] year during the careers of 12 Nobel laureates. However, advantages and disadvantages of the impact factor in comparison with the h-index remain. In addition the author disambiguation problem is more serious for the Author Impact Factor, because all uncited and lowly cited papers have to be taken into account.

The current index which has been proposed recently by Fiala (2014) is equivalent to the timed h-index investigated below. Specifically, Fiala utilized a 3-year window for his so-called h3-index taking into account publications and citations from 3 years. He also proposed alternative variants based on a 2-year publication window and either a 4-year citation window or a sliding 3-year citation window. He further studied another variant applying a 3-year citation window to all previous publications. However, this procedure is in conflict with the spirit of the present approach, because it also takes rather old publications into account.

I have the same reservations against the $h_5$-index determined by Pan & Fortunato (2014) where citations in the current and the last 5 years to all publications are used to determine the index. The authors have already admitted that this index "does not reveal the true current impact of the scientist in those five years." They have indeed also mentioned a so-called incremental h-index for which the considered publications are likewise restricted to the citation window. Unfortunately they did not investigate this incremental h-index. Furthermore, neither Fiala (2014) nor Pan & Fortunato (2014) have studied the dependence of their indices on the length of the utilized citation-time window. Without further analysis, the time windows in Fiala's investigation appear rather short, as can be seen from his results where the index values are small. Similarly, Sidiropoulos, Katsaros, & Manolopoulos (2006) have reduced the impact of older papers in the definition of their contemporary h-index by scaling the total citation frequencies with an inverse power law in dependence on the current age of the papers. But this is a rather elaborate procedure in comparison with taking the naked frequencies, but restricting the evaluated time interval. Below, I present data for different citation times in order to choose a reasonable length of the window.

Due to the definition of the h-index as well as of most of its variants, the index values are restricted to integer numbers so that short time windows and thus resulting small index values reduce the discriminative power of the results as also noted by Pan & Fortunato (2014) for their incremental h-index. Fiala's results reflect this, and he uses the number of publications as a secondary criterion to differentiate between researchers with the same score. The problem can be mitigated if not avoided by utilizing the interpolated version of the h-index (Rousseau, 2006, van Eck & Waldman, 2008, Schreiber, 2008, 2009) leading to a finer distinction.

## 2. The timed index $h_t(y)$

For the following investigations I define the timed h-index $h_t(y)$ for a specific year $y$ taking into account publications in the year $y$ and in the previous $t$ years. Thus this variant is defined like the usual h-index, but restricted to publications in the time window from the year $y - t$ until the year $y$:

$h_t(y)$ is the largest number of papers which have been published by a researcher from the year $y - t$ until the year $y$ and have obtained at least $h_t(y)$ citations each.

This implicitly defines the citation time window: It also covers the year $y$ and the preceeding $t$ years. If $y_0$ indicates the start of the publication activity of a scientist, then for $t = y - y_0$ the index $h_{y-y_0}(y)$ reflects the time evolution of the usual h-index up to the final year of the investigation. In the present case this is the year 2014, for which one obtains the simple h-index $h_{2014-y_0}(2014) = h$. For $t = 2$ the timed index $h_t(y)$ agrees with Fiala's current index h3. It is also equivalent to the incremental h-index mentioned by Pan & Fortunato (2014) if the time window $t$ in my definition of $h_t(y)$ is replaced by their $\Delta t$.

The citation records for the following investigation were determined in the ISI Web of Science database in September 2014 and utilized in my previous investigation (Schreiber, 2014) of $h_r(y)$. The evolution of the timed h-index for my citation records is presented in Fig. 1 for several time intervals $t$.[1] Due to its definition, $h_{t'}(y) \geq h_t(y)$ if $t' > t$, so that the curves for $t'$ can never fall below the curve for $t$: Frequently, the lines in Fig. 1 touch, but they never cross. In the first $t$ years since 1976 the curves for $t$ and $t'$ coincide for $t' > t$, because my first publication dates from $y_0 = 1976$ so that time intervals $t' > t = y - 1976$ do not comprise more publications than the time interval $t$.

For small values of $t$ there are relatively large fluctuations in the evolution of $h_t(y)$. With increasing $t$ these fluctuations are smoothed down and periods of larger and smaller impact can be distinguished. Already for $t = 3$ but more prominently for larger values of $t$ a strong increase can be identified in the early nineties indicating a successful period in my career. This reflects the beginning of a productive phase soon after I got my first professorship and could form a larger research group. Of course, the increase in the curves is not due to the productivity but to the impact in terms of citations to the produced publications. Most curves remain at the level reached in the mid-nineties which means that I maintained my performance. There is an increase for larger intervals $t$ in the late nineties which shows the lasting impact of those publications from the early nineties or in other words those publications kept receiving citations. Another increase of the $h_t(y)$ curves although not as strong as that in the early nineties can be observed from 2006 or 2007 for $t$ years for various values of $t$. This is largely due to the impact of my research in bibliometrics.

---

[1] Here the interpolated h-index is used, which is calculated as $h = c(h)$ from a linear interpolation of the citation frequencies $c(r)$ between ranks $h$ and $h + 1$. This leads to a finer distinction which makes the figure easier to access. The usual h-index can be obtained by truncating the interpolated rational index value to its integer part.

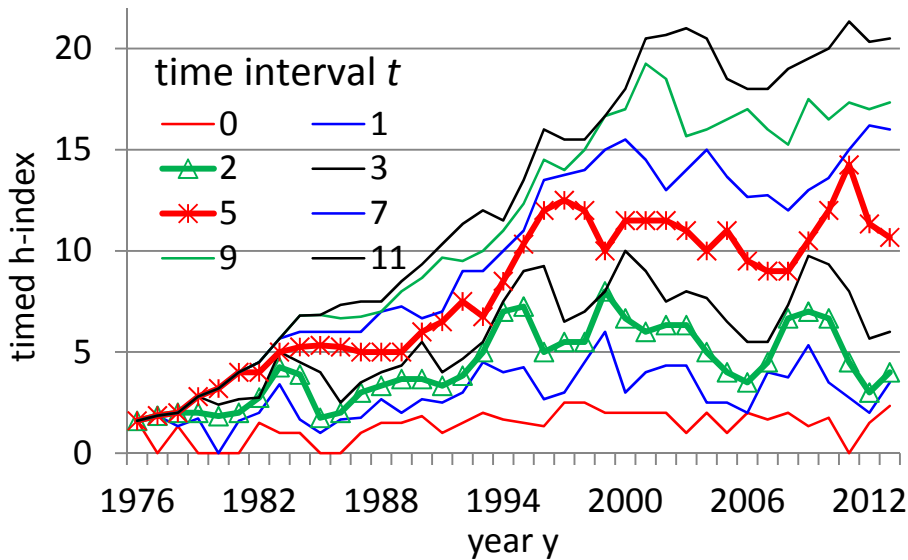

**Fig. 1.** Time evolution of the timed h-index $h_t(y)$ for the publications of the present author in dependence of the length of the time interval $t$.

### 3. The citation time window

The question arises which time interval $t$ would constitute a reasonable citation time window. For this purpose I have determined the number of citations which each paper has received up to the year $t$ after its publication. Expressing these numbers as a percentage of the total citation frequency for each paper allows me to determine e.g. the time interval $t$ which was necessary to obtain half of the total number of citations. Of course, for small numbers of citations these percentages are strongly influenced by a few additional citations. Also for very recent publications the results are not so meaningful, because the number of citations can be expected to grow relatively strongly in a short additional time. Therefore I have first concentrated on the 35 papers which currently contribute to my usual h-index. Among those there are only 5 papers which needed a citation window of $t \geq 10$ years to receive more than half of the total number of citations. 9 papers needed $6 \leq t \leq 9$ years, 19 papers needed $3 \leq t \leq 5$ years. Two papers had reached the median already within two years after publications. Another way round one can say that within 5 years after publication 21 out of the 35 papers had received more than 50% of their total number of citations up to now and 34 had received more than 20%. If one analyzes the 82 papers with at least 20 citations then one finds that 53 papers have received at least 50% of their citations within 5 years after their publication and 78 have received more than 20% in this time interval.

The results of this analysis are displayed in Fig. 2 where the citation time window is displayed in which the lower quartile, the medium, the upper quartile, and the upper decile of the total number of citations for each of these 82 papers were obtained. This visualizes the ageing of the papers in terms of citation frequency. The often short black bars for the lower quartile reflect the short initial period in which 25% of the citations were received. Most of the blue (dark grey) bars are also rather short in correspondence to the above discussion, although there are some outliers in particular for very early papers. The upper decile covers several years in several cases, which means that further citations are coming in slowly.

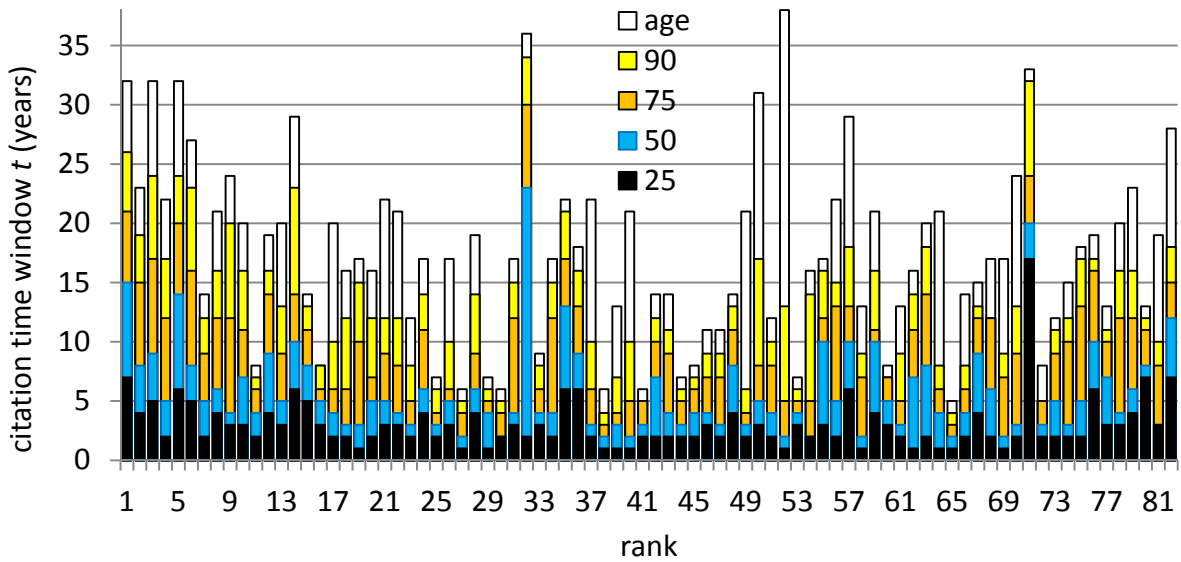

**Fig. 2.** Citation time windows which were necessary to obtain 25%, 50%, 75%, 90% of all citations for each of the 82 papers with at least 20 citations, shown by black, blue (dark grey), orange (medium grey), yellow (light grey) bars (from bottom to top), respectively. The papers are ranked by decreasing citation frequency. The white bars indicate the age of the publication. In most cases this is (nearly) equal to the citation time window $t$ in which 100% of the citations were received: 66 of these 82 papers were cited recently, namely in 2013 or 2014. Note that the citation time window includes the publication year, so that the factual length of the window is between $t$ and $t + 1$ years, depending on the month of publication.

In order to obtain a more comprehensive picture, I have grouped the publications in such a way that the total number of citations to papers from each group is approximately equal to 800 citations, i.e. about 15% of all the 5255 citations that my papers have received. (The last group comprises the remaining, namely only about 10% of all citations.) One can see that the median is reached in most groups for $t \leq 5$ years. I find it noteworthy that the group of most cited papers reaches higher percentages much more slowly. This corresponds to the above observation: old and highly cited publications keep receiving citations for a long time. It is in agreement with previous findings (Wang, 2013 and references therein) "that highly cited papers had a slower ageing process". On the other hand, for the last group which comprises the lowly cited papers with up to 9 citations the respective curve in Fig. 3 rises fastest, indicating that these few citations have mostly been obtained rather soon after the publication. This effect is enhanced, because most recent papers fall into this group.

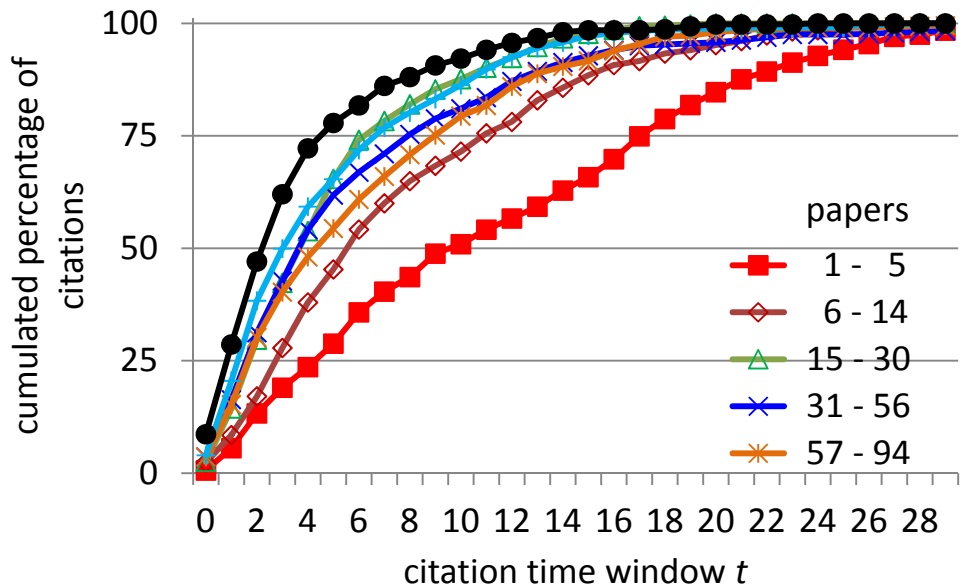

**Fig. 3.** Percentage of citations which were received in the publication year and the subsequent $t$ years. Here the papers are grouped in such a way that all papers in a group received about 800 citations altogether up to 2014. The ranks for each group are listed in the inset.

If one displays the absolute number of citations in the year *t* after the publication year for the same groups, then the maximum is reached already for *t* = 2 in 5 of the 7 groups, see Fig. 4. For the group with the lowly cited papers the maximum occurs already for *t* = 1 in agreement with the above observations. The maximum for the highly cited papers can also be found for *t* = 2 years but it is the lowest maximum in this plot and there are several further low local maxima; for large values of *t* this curve is much higher than all the other curves. This again reflects the long lasting impact of the publications and the slower ageing.

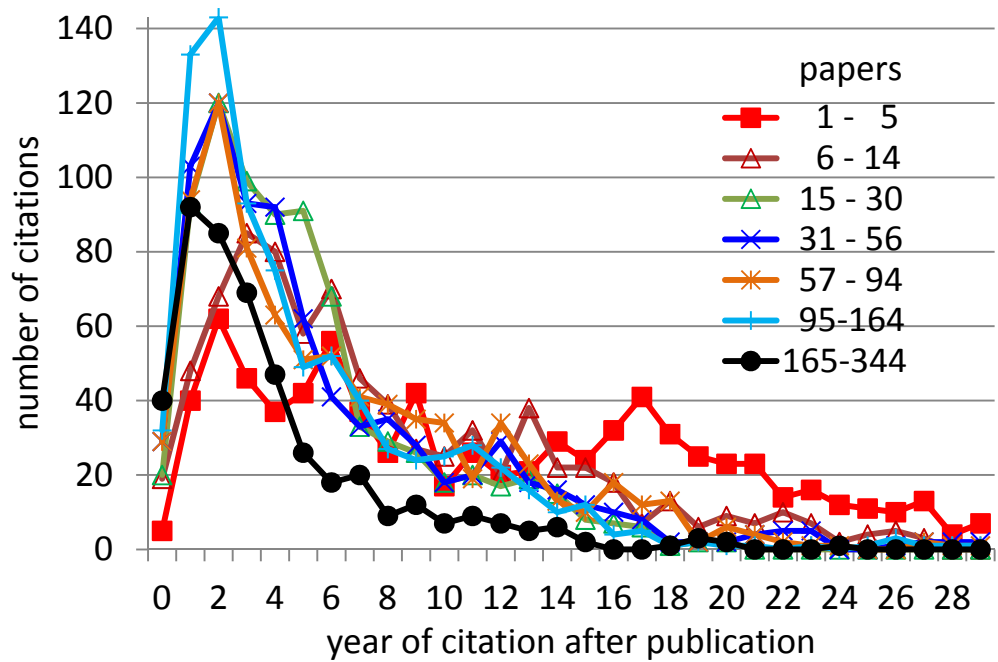

**Fig. 4.** Number of citations received in the *t*-th year since publication. Here the zeroth year is not a full year because it means the year of publication.

In conclusion, *t* = 2 as used by Fiala (2014) appears to me as somewhat too small, because it just includes the maximum of most of the curves in Fig. 4. But *t* = 5 as suggested by Pan & Fortunato (2014) is probably a reasonable choice. It is also practical, because the values of $h_5(y)$ are not so small that fluctuations are dominant. On the other hand *t* = 5 is still reasonably short so that the index values really indicate the impact of the recent performance and are not dominated by citations to rather old publications.

Rousseau & Ye (2008) have proposed a dynamic h-type index based on the increase of the h-index. That variant takes all papers into account, but can be restricted to the increase of the index in recent years. Without further evidence, the authors "claim that a period of 10 or 5 years … is appropriate" for hiring purposes. Wang (2013) has studied the correlation of citation counts in different years with the total number of citations after 31 years. The analysis shows that different time windows are appropriate for citation impact predictions in different fields and also for screening out elite papers and elite scientists. Bornmann, Leydesdorff, & Wang (2014) have shown that for short citation time windows up to 3 years such predictions can be significantly improved by taking further variables into consideration, like the journal impact factor, the number of cited references, and the length of the paper. However, as this procedure needs much more data, its application is not so practical in evaluations. Therefore, I refer to Wang's (2013) data and note that in physics a citation time window *t* = 5 (i.e. 6 years including the publication year) yields a Spearman rank correlation of 0.898, which corroborates my above choice.

### 4. Summary

The investigated *timed* h-index allows one to evaluate the performance of a researcher in terms of received citations in a *timely* way concentrating on recent *topical* publications and citations. It can be easily obtained from the Web of Science by restricting the publication time window in the search and then adding the yearly citations up to the year *y*. It is particularly instructive to view the evolution of the timed h-index during the entire career in order to see whether the evaluated scientist is still successfully publishing or whether relatively high and possibly strongly increasing values of the usual h-index are due to old publications and thus to passée achievements.

Further investigations are necessary to confirm that the choice of $t = 5$ years yields indeed a reasonable length of the publication time and citation time window. It would also be interesting to see whether the timed h-index is a better choice than the usual h-index, if one tries to predict scientific success as Acuna, Allesina, & Kording (2012) and Mazloumian (2012) have done. Certainly, one major criticism against the predictability of the h-index, namely that the order restriction (*h* cannot decline) makes conventional significance tests not meaningful (García-Pérez & Núñez-Antón, 2013) does not apply to the timed h-index. Another problem with predicting the h-index, namely that one should better distinguish different career-age cohorts (Penner, Petersen, Pan, & Fortunato, 2013) does not occur for the timed h-index: The timed h-index does not favor older scientists, because unlike the usual h-index it does not count the citations to old publications. Nevertheless, older scientists may still have an advantage, because they have been visible for a longer time which might influence other researchers to cite them, not necessary their old but also new papers, more frequently. A similar argument applies to famous researchers.

**References**

**Acuna, D.E., Allesina, S., & Kording, K.P. (2012).** Predicting scientific success. *Nature*, 489, 201-202.

**Bornmann, L., Leydesdorff, L., & Wang, J. (2014).** How to improve the prediction based on citation impact percentiles for years shortly after the publication date? *Journal of Informetrics*, 8, 175-180.

**Fiala, D. (2014).** Current index: A proposal for a dynamic rating system for researchers. *Journal of the Association for Information Science and Technology*, 65(4), 850-855.

**García-Pérez, M. A. & Núñez-Antón, V. (2013).** Correlation between variables subject to an order restriction, with application to scientometric indices. *Journal of Informetrics*, 7, 542–554.

**Hirsch, J.E. (2005).** An index to quantify an individual's scientific research output. *Proceedings of the National Academy of Sciences*, 102(46), 16569-16572.

**Hirsch, J.E. (2007).** Does the h-index have predictive power? *Proceedings of the National Academy of Sciences*, 104(49), 19193-19198.

**Mazloumian, A. (2012).** Predicting scholars' scientific impact. *PLoS ONE*, 7, e49246.

**Pan, R.K. & Fortunato, S. (2014).** Author Impact Factor: Tracking the dynamics of individual scientific impact. *Scientific Reports*, 4, 4880.

**Penner, O., Petersen, A. M., Pan, R. K., & Fortunato, S. (2013).** The case for caution in predicting scientists' future impact. *Physics Today*, 66, 8–9.

**Rousseau, R. (2006).** Simple models and the corresponding h- and g-index. http://eprints.rclis.org/7501

**Rousseau, R. & Ye, F.Y. (2008).** A proposal for a dynamic h-type index. *Journal of the American Society for Information Science and Technology,* 59(11), 1853-1855.

**Schreiber, M. (2008).** An empirical investigation of the g-index for 26 physicists in comparison with the h-index, the A-index, and the R-index. *Journal of the American Society for Information Science and Technology*, 59(9), 1513-1522.

**Schreiber, M. (2009).** The influence of self-citation corrections and the fractionalised counting of multi-authored manuscripts on the Hirsch index. *Annalen der Physik* (Berlin) 18, 607-621.

**Schreiber, M. (2013).** How relevant is the predictive power of the h-index? As case study of the time-dependent Hirsch index. *Journal of Informetrics*, 7, 325-329.

**Schreiber, M. (2014).** A variant of the h-index to measure recent performance. *Journal of the American Society for Information Science and Technology*, in print, *arXiv:*1409.3379v2.

**Sidiropoulos, A., Katsaros, C., & Manolopoulos, Y. (2006).** Generalized h-index for disclosing latent facts in citation networks. *Scientometrics*, 2, 253-280.

**Van Eck, N.J. & Waltman, L. (2008).** Generalizing the h- and g-indices. *Journal of Informetrics*, 2(4), 263-271.

**Wang, J. (2013).** Citation time window choice for research impact evaluation. *Scientometrics* 94, 851-872.